# What We Risk Losing When Creating Gets Easy

## FRICTION, JUDGMENT, AND CRITICAL REFLECTIVE PRACTICE WITH GENERATIVE AI IN CREATIVE WORK

**Shehryar Saharan [1,2], Shehroze Saharan [3], Roxanne Ziman [4], Nicholas Woolridge [1], and Gaël McGill [5]**

**ABSTRACT** Generative AI (GenAI) in creative practice can help narrow the gap between intention and output, but in so doing changes the very nature of that creative process. In this position paper, we argue that the friction of making is not overhead to be removed, but essential to creative work: the resistance through which judgment is built and refined. Rejecting both outright refusal and uncritical adoption, we call for *critical reflective practice*: the deliberate, ongoing, and situated weighing of when to use or refuse GenAI in creative work, treating the formation of judgment as an epistemic virtue that design and pedagogy should (continue to) uphold. Two voices, the GenAI *Skeptic* and GenAI *Enthusiast*, drawn from our professional and personal experiences, argue with each other and with us throughout. We close with open questions for researchers, educators, and practitioners navigating the grey areas of GenAI in creative practice.

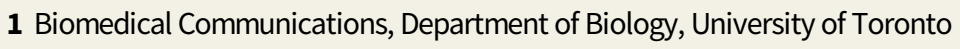

**1** Biomedical Communications, Department of Biology, University of Toronto

**2** College of Engineering, University of Guelph

**3** Office of the Vice-President, Digital Transformation & Chief Information Officer, George Brown Polytechnic

**4** Department of Informatics, University of Bergen

**5** Center for Molecular and Cellular Dynamics, Department of Biological Chemistry and Molecular Pharmacology, Harvard Medical School

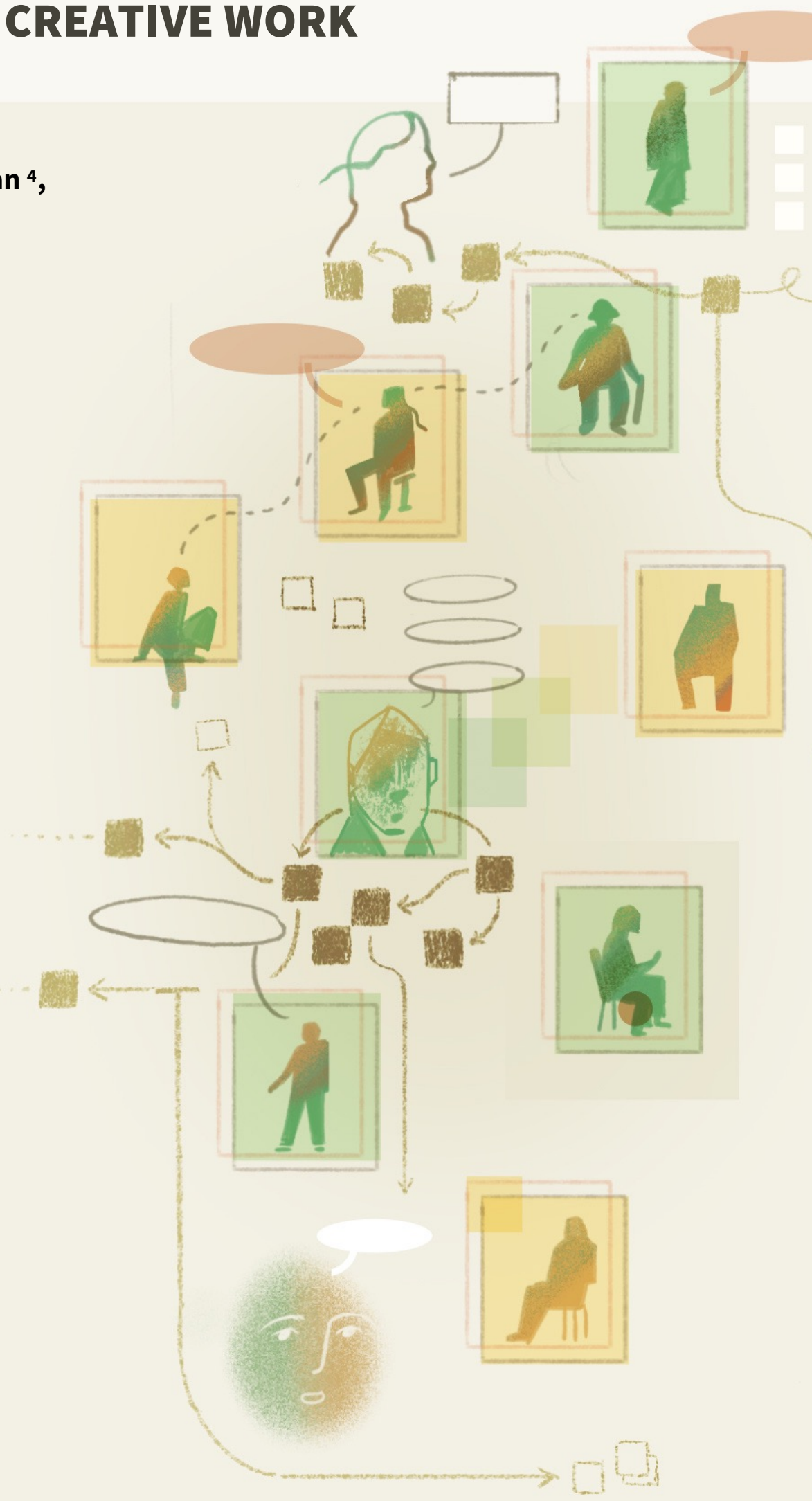

**Two voices appear in the margins of this paper:** the GenAI **Skeptic** and GenAI **Enthusiast**, personas whose voices are drawn from our earlier empirical work (Ziman et al. 2026), as well as our own practice and teaching. In many ways, they reflect our own experiences in holding these tensions, often simultaneously, through the process of creating. Throughout this work, these interlocutors engage with one another and with us from opposing perspectives: each is given license to overstate, push back, and refuse the resolutions we are tempted to offer too quickly. These personas are also how we register the visceral dimension of this debate which we have directly observed and felt. They also help surface the '*black-and-white thinking*' that we are trying to do away with, toward **critical reflective practice** (*see page 4*).

## BENEFITS, LIMITS, AND WHAT REMAINS UNCLEAR

There is a real case to be made for GenAI in creative practice. Across multiple creative domains, practitioners report that current tools can, in some workflows, help narrow the gap between what they can imagine and what they can produce. As a new generation of creative support tools (CSTs) (Palani & Ramos, 2024), GenAI can make some forms of exploration faster, lower the cost of experimentation, and shift the practitioner's efforts from certain forms of slow production toward creative decisions involving curation and refinement.

GenAI tools can also speed up the generation of alternatives and suggest creative directions in ways that open up earlier and wider exploration (Palani & Ramos, 2024; Chandrasekera et al., 2024). In several empirical works involving GenAI use, participants have reported higher creativity, usefulness, novelty, or overall design quality at the level of individual outputs, though these effects vary by task, domain, and user group (Doshi & Hauser, 2024; Chandrasekera et al., 2024; Fu et al., 2024; Holzner et al., 2025). The gains appear especially visible for some novices or less GenAI-experienced creators.

> ***The Enthusiast:*** *Yes! GenAI has the potential to lower the barrier to creative work and make production feel possible for more people. Individuals who have never felt confident in their technical or artistic skills can now move from an intention to a finished work in far less time and effort.*

We gain ease and efficiency when we use GenAI tools to brainstorm, perform tedious technical tasks, and move quickly beyond the blank page to take our initial, sometimes vague ideas to a polished, finished product in much less time, and, perhaps for some, with less technical acuity than previously required (Palani & Ramos, 2024).

But these stages of the creative workflow carry meaningful *friction*: the early uncertainty before a project has taken shape, the slow articulation of intent, the sometimes arduous work of technical execution.

***The Skeptic:*** *Agreed - the discomfort of the blank page is where designers figure out what they actually want to say, and how they want to say it. Friction matters in learning, in experimenting, through making...*

**What do we risk losing, then, when we reduce friction in creative work? What do we risk losing when creating gets 'easy'?** More precisely, what happens when tools reduce or relocate the moments of resistance through which practitioners typically develop *judgment*?

**A moment to clarify what we mean by *judgment***: it is the situated, largely tacit capacity through which practitioners discern what a work demands, practically, ethically, empathetically, and aesthetically, where the answer cannot be derived from rules or external evaluation alone. In this sense, judgment is a reflective capacity formed through sustained engagement with the resistances of material, medium, constraint, embodied practice, and context (Schön, 1983; Sennett, 2008; Nelson & Stolterman, 2012).

> ***Insight*** *is deeply entangled with judgment, and worth naming here as well. Practitioners think through their making, discovering what they understand by working ideas out in material form (Ingold, 2013). Insight is what surfaces in this process: the often unanticipated understanding of a subject that emerges when complex material is wrestled into a form that holds. For instance, the act of representing data, structures, or processes can synthesize what the practitioner knows and externalize their mental models, surfacing connections that may not have been evident before the work began (McGill, 2022).*

This process takes time and conscientious effort. We argue that when tools absorb or remove friction before judgment has formed, some creative workflows risk shifting from open-ended authoring toward the orchestration, selection, evaluation, and refinement of outputs that already embody a visual or structural logic. Although not everyone has taken up this practice, it is increasingly visible in GenAI-adapted creative workflows, where practitioners describe their roles less as executing each element of a work and more as orchestrating information, tasks, models, and outputs (Palani & Ramos, 2024; Tsao et al., 2025). It is also visible in emerging GenAI CSTs, where design attention is increasingly directed toward mixed-initiative interaction, sketch-based steering, and other ways of guiding, combining, and refining generated alternatives (Lin et al., 2023; Zhang et al., 2023).

**We come to these tensions and concerns from a particular vantage point:** as practitioners, educators, and researchers working across visual communication, design education, and the integration of emerging technologies into contexts where creative practice and learning are deeply entangled. **We have come to recognize judgment as something built slowly, through years of contact with the 'friction of making'; we are now encountering a shift in which GenAI tools reduce, redistribute, or absorb the very friction-full aspects of the creative workflow through which that judgment has traditionally formed.** These tools are challenging our notion of what creative work is and how we as practitioners understand our own craft (Shelby et al., 2024).

***The Skeptic:*** *The formation of judgment takes shape through friction. Take away the friction early enough, and you do not (always) get a faster path to the same place.*

***The Enthusiast:*** *The destination practitioners arrived at before GenAI tools existed was shaped by the constraints of preceding tools too.*

**The friction of making is not overhead. It is part of the medium of creative work, and the loss most worth thinking about is the process by which judgment takes shape.** At their best, creative tools do not remove decision-making steps inherent to the creative process; they '*get out of the way*' and facilitate the path by which we engage with the medium at hand. That engagement is the heart of the work. We argue that it is worth protecting.

## FRICTION AS CONSTITUTIVE

By friction, we mean the resistance of the work itself: the way an approach reveals its limits halfway through, and the way an idea has to be wrestled into a form that holds (Schön, 1983; Sennett, 2008; Nelson & Stolterman, 2012). Picture a designer sketching a layout. They commit a line. They try a composition. They redraw. They try something unfamiliar, and in the not-quite-working, the work reveals something they had not understood about the problem. Every mark is both a commitment and a question, a provocation, and an inquiry into what the artifact may become. Across enough marks, the eye learns to discriminate, which lines work, which do not, which marks suit which contexts; that learning is what makes the practitioner's judgment reliable**.** That is the kind of friction we are concerned with: friction is constitutive of the work, not incidental to it. It is also the form of friction that most accessible GenAI tools most readily consume. The argument that GenAI expands the range of options a practitioner considers is true in one sense, but the breadth of pre-formed options is not the same as dedicated and disciplined inquiry. The practitioner risks locking in sooner, not later: the alternatives arrive already resolved, and the friction-full work of figuring out what one is trying to craft becomes offloaded to the tool.

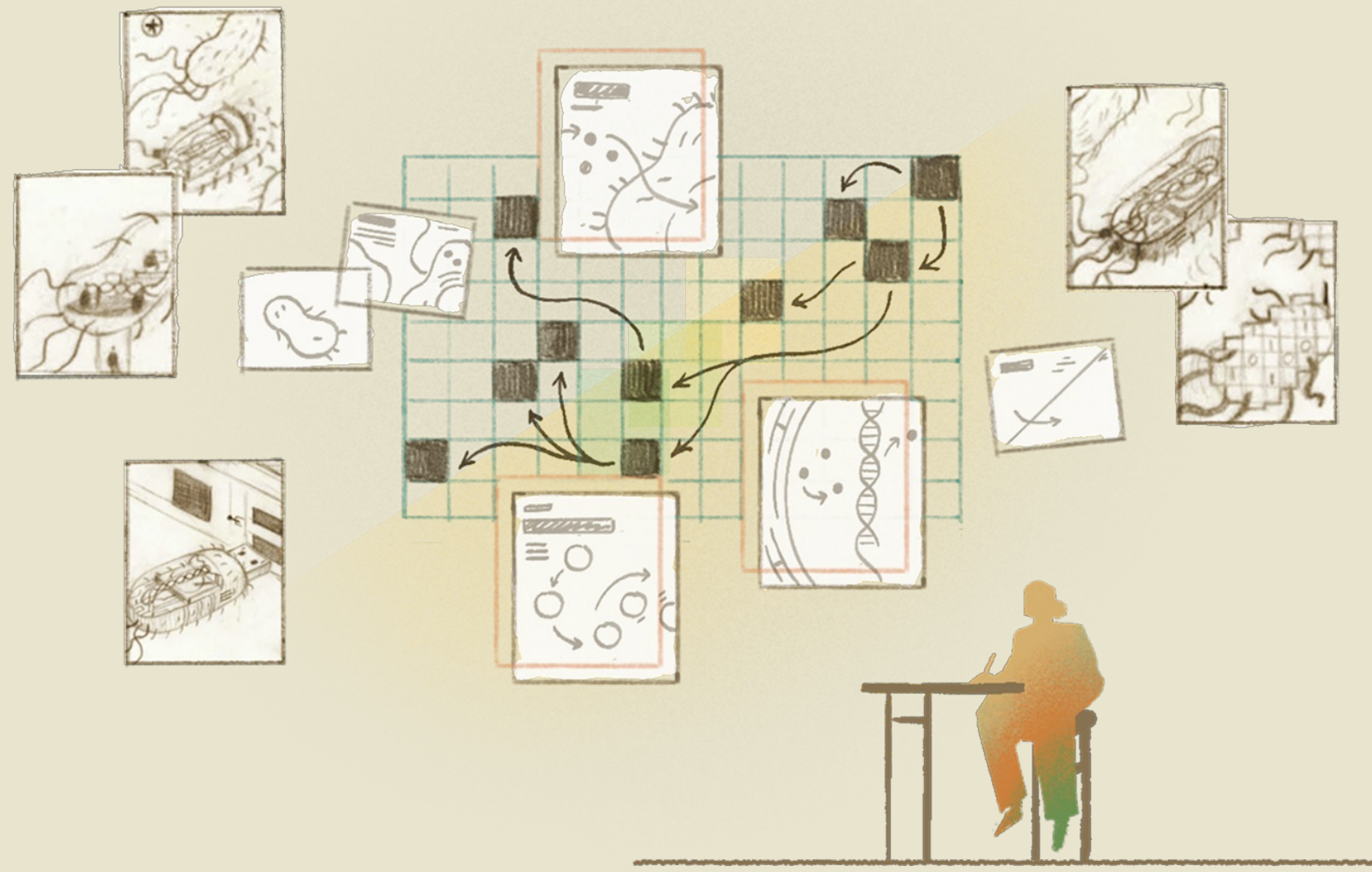

Schön's reflection-in-action runs in the same direction (Schön, 1983). It is a refined capacity to be surprised by the work, to notice when it resists, and to learn from that resistance. Recent work in creative coding pedagogy has documented this dynamic empirically: when tools are "*too fast and seamless,*" students may "*only be able to express directorial agency rather than learned craft,*" reducing access to the kinds of friction through which judgment is afforded and formed (McNutt et al., 2025). For McNutt and colleagues, the issue is not speed itself, but the loss of pedagogically useful friction; decreased resistance may also remove opportunities for reflection, understanding, honing technical skill, and the cultivation of craft.

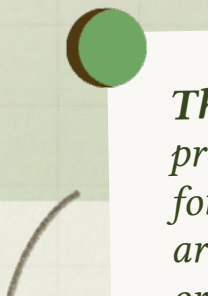

***The Enthusiast:*** *There is resistance in working with GenAI tools too. A prompt may produce something that is not quite right, and the process can force you to clarify your intentions, though not all creative intent is articulable. That resistance is not always the same as working through form or material directly, but it can still become part of the design process.*

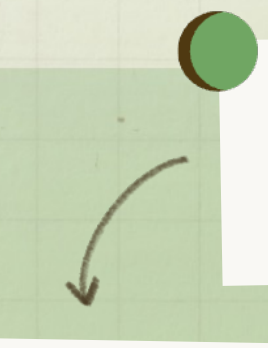

***The Skeptic:*** *Fair. But there is a difference between resistance that teaches you something about the work and resistance that teaches you how to manage the system. Both can build judgment, but not necessarily the same judgment. Both are real, but they are not interchangeable.*

***The Enthusiast:*** *That may not hold up in practice. Figuring out how to prompt for something forces you to articulate what you actually want, and that articulation is often where the thinking happens.*

***The Skeptic:*** *Then it is worth asking: does it matter in what medium that articulation happens? When you prompt, you articulate in words, outside the medium the work will ultimately live in. When you sketch, you use similar materials, marks, and gestures; the thinking happens through the medium, not apart from it.*

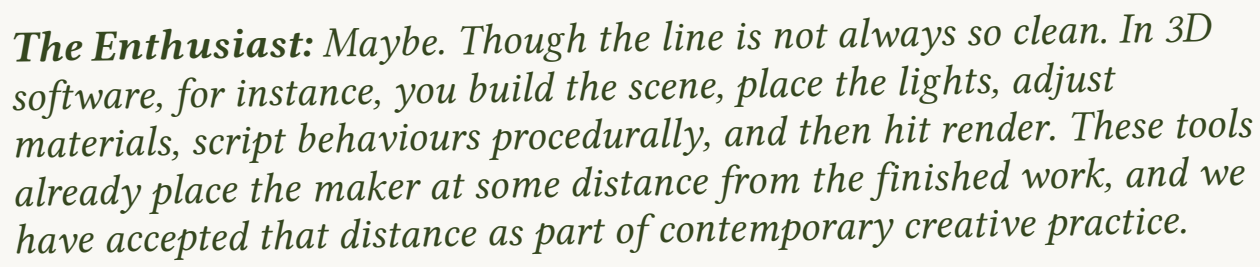

***The Enthusiast:*** *Maybe. Though the line is not always so clean. In 3D software, for instance, you build the scene, place the lights, adjust materials, script behaviours procedurally, and then hit render. These tools already place the maker at some distance from the finished work, and we have accepted that distance as part of contemporary creative practice.*

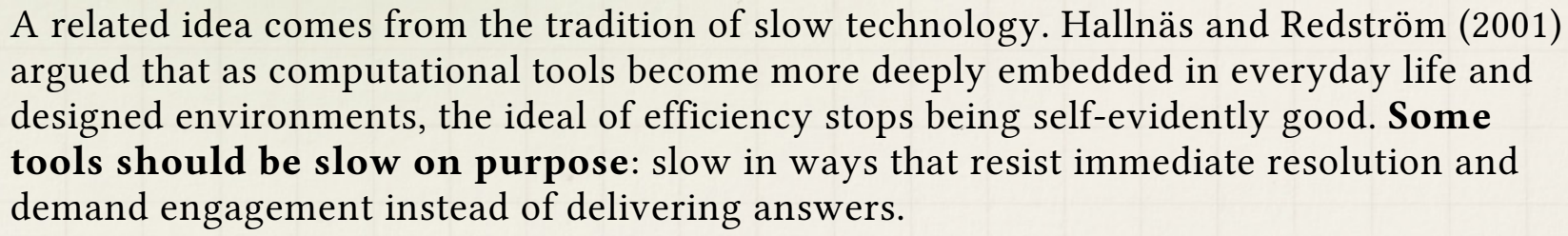

A related idea comes from the tradition of slow technology. Hallnäs and Redström (2001) argued that as computational tools become more deeply embedded in everyday life and designed environments, the ideal of efficiency stops being self-evidently good. **Some tools should be slow on purpose**: slow in ways that resist immediate resolution and demand engagement instead of delivering answers.

Van der Burg et al. (2026) bring this tradition into AI in design education, arguing that current paradigms of GenAI use prioritize fluency and immediate output in ways that discourage reflective engagement. They propose instead that designing for slowness and productive encounters with AI's underlying components and limitations can recover the kinds of critical and constructive engagement that fast, output-oriented tools tend to displace. Weisz et al. (2024) similarly address the risk of overreliance in their design principles for GenAI, recommending that responsible systems "*use friction to avoid overreliance*" by incorporating mechanisms that intentionally slow practitioners down at key decision points, prompting reflection before action.

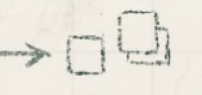
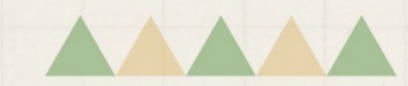
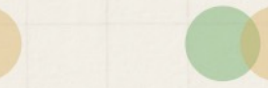
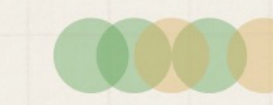

Now, picture the same designer asking a generative model to produce a layout. In the vast majority of widely adopted GenAI tools and workflows, the system produces outputs that are already partially or fully resolved. The output may be excellent, but the sites of friction have been shortened (if not bypassed). What disappears, or is at least compressed, is a sequence of small decisions through which the designer would have come to understand the material, the problem, and their own intentions with and toward each (Chiang, 2024). **The practitioner reaches a destination, but perhaps a different one entirely.**

**So, how much of that destination can the practitioner claim as their own?** Chiang (2024) suggests that the very feature that makes GenAI tools appealing (*minimal input and substantial output*) is also what attenuates the practitioner's claim on the result. This framing has a productive design corollary: if thin input produces thin authorial presence, then the response is to design tools that elicit richer intent from users, a challenge Kreminski (2025) terms *"lensing the imagination."* When a brief prompt produces a fully resolved work, most of the choices were not made by the practitioner, but inherited, on average, from prior work the system was trained on. Prompting and refining are different from the work of building an artifact, and they produce a different relationship between the maker and the output.

This shift in role carries implications. For instance, McGuire et al. (2024) found that individuals are less creative when revising an AI-generated draft than when creating content themselves, unless the interaction is framed as co-creation. The role the practitioner thinks they are playing matters as much as what they are actually doing. This effect is not confined to late-stage editing. Wadinambiarachchi et al. (2024) show that early contact with AI outputs can fix practitioners on the first plausible idea and narrow the divergent thinking that exploration depends on.

***The Enthusiast:*** *This describes a very particular kind of AI use, and perhaps, not the most thoughtful kind. A designer working seriously with these GenAI tools is iterating, rejecting, revising, and prompting again. Treating the first generation as the whole interaction is a flat description.*

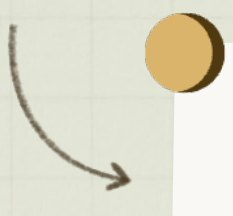

***The Skeptic:*** *Iterating on a generated output is not the same as creating work without GenAI, however many times you do it. The iteration happens in response to something already formed and committed to a logic.*

The Enthusiast and Skeptic are both right that the picture above is incomplete. Translating intent into a form a GenAI system can act on is one kind of thinking, one that may involve iteration, evaluation, redirection, and gradual refinement of intent through repeated contact with the model's outputs (Palani & Ramos, 2024; Zhou & Lee, 2024; Lin et al., 2023; Alharthi, 2025). Wrestling with material until it yields a form is another. The two are not necessarily opposed, but they are not interchangeable either.

## TOWARD A CRITICAL REFLECTIVE PRACTICE

**Ultimately, critical reflective practice is inseparable from the question of how we adopt and integrate GenAI into creative workflows.** If the loss worth attending to is the loss of the process, the response is not to refuse the tools or to embrace them uncritically. Adoption is not inevitable, and refusal is not impossible**. Our response is to take their integration seriously as a design problem and a pedagogical problem in its own right.** Without a culture of critical reflection on how AI tools are taught, adopted, and designed, we risk normalizing creative workflows and works that are increasingly monotonous, uncritical, and compliant.

By **critical reflective practice**, we mean the deliberate, informed, ongoing, and situated weighing of when to use and when to refuse GenAI in creative work; the term echoes Agre's (1997) "*critical technical practice,*" a call for AI researchers to bring humanistic scrutiny to their own technical assumptions. Working within this grey middle ground, we recognize that the same tool may support judgment in one moment and diminish it in another. A *critical* approach interrogates the values built into these tools: what they afford, what they allow us to construct, and how they shape and limit the kinds of judgment that can be exercised through them. A *reflective* approach turns the same scrutiny inward, asking how our own creative practice changes when we work with these tools, and what we are becoming as practitioners through that work.

The friction that supports the formation of judgment should not be sacrificed in the name of ease or speed. We treat this as both a design and pedagogical commitment, one to be integrated into our CSTs and teaching practices.

***The Enthusiast:*** *The question is not whether a tool is fast or slow, but whether it helps me stay with the decisions that matter. A good tool can accelerate without making me absent from the work.*

***The Skeptic:*** *But many tools do not ask which decisions matter. They make fluency the default and reflection the exception. The design question is not just what the tool allows, but what it rewards.*

Supporting this view, Van der Burg et al. (2026) define *Reflective AI* as "*a slow technology approach to AI in creative practice that treats AI systems as mediums for self-reflection rather than generators of design outputs.*" Their work offers a tool-side response: AI systems can be designed to afford reflection by exposing and slowing engagement with the processes through which outputs are produced. McGuire et al. (2024) sharpen the practitioner-side counterpart, arguing that "*people must occupy the role of a co-creator, not an editor, to reap the benefits of generative AI in the production of creative works.*"

Together, these works gesture toward a tool-side commitment (to systems that encourage reflection rather than simply delivering outputs), and a practitioner-side commitment (to occupying a role that allows judgment to form). We extend both: critical reflective practice asks not only how tools should be designed and what role practitioners should occupy, but what conditions across tool design, pedagogy, and individual practice make the formation of judgment possible, and which choices, deliberately or by default, shut it down.

This commitment generates questions at many scales: tool design, pedagogy, individual practice, etc.... We use the remainder of this section to lay some of them out as a set of provocations.

*To stand behind a work has often meant understanding how it came to be. Which parts of this work do practitioners need to struggle through in order to understand, revise, and claim it as their own?*

*Early outputs shape the trajectory of the work; how different would the outcome be if those initial suggestions were absent?*

*How do we teach students to distinguish between using GenAI to extend judgment and using it to substitute for judgment that has not yet formed?*

*How do GenAI tools shape the creative processes, roles, and division of labour within the creative industry?*

*If students are able to bypass friction, what exactly are they learning to do? What remains unlearned?*

*Are tools designed to surface the moments where a practitioner's judgment is being shaped, or to reduce friction in service of ease and efficiency alone?*

*How might tools make visible the difference between what the practitioner decided, what the system generated, and what emerged through interaction between the two?*

*How do we cultivate critical reflective practice as a shared habit across tool design, pedagogy, and practice?*

### WE CLOSE WITH ONE FINAL EXCHANGE

***The Skeptic:*** *My worry is not really about the finished product. It is about everything we move through to get there. The act of making carries its own kind of knowing: cognitive, emotional, and even physical. Through the process of making, we grow. We gain insight, not only about the thing we are working to create, but about ourselves as well.*

***The Enthusiast:*** *I will not pretend that GenAI tools cannot shorten that process. They can. But they do not have to make the maker absent from it. A good tool can accelerate some parts of the work while still leaving room for the decisions, hesitations, and discoveries that matter. The question is not whether the path becomes shorter, but whether the practitioner is still changed by moving through it.*

The friction of making is not the only thing at stake when GenAI enters creative practice. But we think it is among the easiest and most valuable things to lose as creators. Critical reflective practice, at its most basic, asks us to keep watch.